\documentclass[manuscript]{aastex}
\usepackage{graphicx}
\usepackage{epstopdf}

\shorttitle{Calculator for Supernova Remnants}
\shortauthors{Leahy \& Williams}

\begin{document}


\title{A Python Calculator for Supernova Remnant Evolution}


\author{D.A. Leahy and J.E. Williams}
\affil{Department of Physics $\&$ Astronomy, University of Calgary, Calgary,
Alberta T2N 1N4, Canada}



\begin{abstract}
{A freely available Python code for modelling SNR evolution has been created. 
This software is intended for two purposes: to understand SNR evolution; and to use in modelling observations of SNR for obtaining good estimates of SNR properties.
It includes all phases for the standard path of evolution for spherically symmetric SNRs. In addition, alternate evolutionary models 
are available, including evolution in a cloudy ISM, the fractional energy 
loss model, and evolution in a hot low-density ISM.
The graphical interface takes in various parameters and produces 
outputs such as shock radius and velocity vs. time, SNR surface brightness 
profile and spectrum.
Some interesting properties of SNR evolution are demonstrated using the program. 
}

\end{abstract}





\section{Introduction}

The study of supernova remnants (SNRs) is of great interest in astrophysics
(see \cite{2012Vink} and references therein for a recent review). 
SNRs provide valuable information relevant to stellar evolution, the evolution of the Galaxy and its interstellar medium. 
SNRs are the dominant source of kinetic energy input into the interstellar medium.

SNRs are observed primarily in X-rays, by emission from hot interior gas with 
temperature $\sim$1 keV, and in radio, by synchrotron emission from 
relativistic electrons accelerated by the SNR shockwave. 
The observational constraints for different SNRs are often different in nature. 
They depend on the brightness of emissions in different wavebands by a given 
SNR and by the instruments used to observe that SNR.
Only a small fraction of the $\sim$300 observed SNRs in our Galaxy have been well 
enough characterized to determine their evolutionary state, including explosion 
type, explosion energy and age. 
In order to expedite characterization of SNRs, we present a set of SNR models and a software implementation in Python. 

The set of models consists of a wide set of models constructed previously by other authors (details given below). 
We carry out the additional step of consistently joining different stages of evolution, which in many cases has not been done before. 
The resulting software facilitates the process of using different constraints 
from observations to obtain SNR physical properties of interest.

The structure of the paper is as follows.
Section 2 presents an overview of supernova remnant evolution. 
Section 3.1 describes some of the overall properties related to calculation
of SNR models, such as characteristic time and radius scales and electron-ion
temperature equilibration.   
Section 3.2 describes the calculations for the standard evolutionary path of a SNR.
Section 3.3 describes alternate evolutionary paths, including cases of SNR in
a cloudy ISM, the fractional energy loss model for SNR, and SNR evolution in
a hot low-density ISM.
Section 3.4 briefly describes the software.
Section 4 gives results from example calculations done with the software
and Section 5 gives a brief summary.

\section{Overview of Supernova Remnant Evolution}

A supernova (SN) explosion creates a SNR starting with the ejection of the SN
progenitor envelope at high speed, typically 5000-10000 km/s.
General descriptions of SNR evolution are given in numerous places (e.g. see \cite{1988cioffi} and \cite{1999truelove}, hereafter referred to as TM99).
In the case where a compact remnant is formed, the envelope is the progenitor less the core that becomes the remnant.  
In the case of no remnant, such as for thermonuclear Type Ia explosions, the ejected mass 
(envelope) is the entire SN progenitor. 
The high-speed ejecta interacts with the surrounding gas (circumstellar medium, CSM,
or interstellar medium, ISM) creating a shock front at its outer edge.
The CSM is the nearby surrounding gas affected by the progenitor star, whereas
the ISM is more generally the large-scale gas between the stars. 
The general picture of SNR evolution is an initial ejecta-dominated (ED) stage, in which 
the effect of the ejected mass and explosion energy are both important. 
This evolves into a Sedov-Taylor (ST) phase, where only the explosion energy remains 
important because the mass swept up by the SN shock exceeds the ejected mass. 
For ED and ST phases, radiative energy losses by the hot shocked interior are negligible. 
As the SNR becomes older, radiative losses become more important until the 
post-shock pressure becomes significantly reduced. 
This causes formation of a dense cool shell just behind the shock front.
However, the hot low density interior is unaffected by energy losses so the SNR is
driven by the interior pressure acting on the dense shell.
The SNR is then in the pressure driven snowplow (PDS) phase. 
During the entire evolution the outer shock front is slowing down with time. 
During the PDS phase, it can slow down enough that it is not distinguishable from random
motions in the ISM. 
This is the termination (or merger) of the SNR with the ISM.
In some (rare) cases, the SNR can lose its interior pressure before merger, resulting
in an additional phase, where the dense shell coasts outward, conserving momentum.
This phase is called the momentum-conserving shell (MCS) stage.  

The unshocked part of the ejected envelope has a homologous velocity profile ($v \propto r$ at fixed time). 
The outer edge interacts in the form of a shock structure with the CSM or ISM (hereafter, we use the term CSM, but whether the 
surrounding gas is CSM or ISM depends on the particular situation of the progenitor and distance of
the shock from the SN).
From outside to inside, this shock structure consists of:  the undisturbed CSM, 
the outer shock moving into the CSM, a layer of shocked CSM, 
the contact discontinuity separating the shocked CSM from the shocked envelope, 
the layer of shocked envelope, the inner (or reverse) shock moving inward relative to the ejecta,
and the undisturbed ejecta.
After the reverse shock reaches the center of the SNR, the entire ejecta is fully shocked.
Reflected shocks and sound waves are generated at this time, and die out over time 
(e.g. see CMB88). These are perturbations on the overall structure and are ignored in
the models here (and the analytic models in CMB88 and TM99).

We assume that the structure is spherically symmetric so that it depends only and radius 
and time.
The pre-shock external medium (CSM) is assumed to have constant density or a $1/r^2$ density 
profile centered on the SN. 
The latter is appropriate for a stellar wind with steady mass loss rate.
Thus the CSM density is a power-law $\rho_{CSM}=\rho_s r^{-s}$, with s=0 (constant
density medium) or s=2 (stellar wind density profile).
The unshocked ejecta is taken to have a power-law density profile $\rho_{ej}\propto r^{-n}$.

With these assumptions the evolution of the forward and reverse shock has been
calculated for the ED through ST phases (TM99). The early part of the ED phase 
follows a self-similar evolution, calculated by \cite{1982Chev}.
We used the TM99 analytic solutions for our modelling software, with some additional
features as described below.
There is no smooth transition from the s=2 ED and ST phases to the later radiative phases
(e.g. TM99).
Thus we consider the later phases only for the s=0 case.
For the radiative phases we follow the treatment of  CMB88, but modify their solutions
to be consistent with the earlier ED to ST evolution.

We also consider some alternate models for SNR evolution which include
additional physical effects.
In particular, we include models for a SNR in cloudy ISM of \cite{1991WL} (hereafter  WL91), 
the constant energy loss model of \cite{2000LK} (hereafter LK2000), and the
model for SNR in hot ISM of \cite{2005TW} (hereafter TW2005).

\section{Model calculations}

\subsection{General discussion}

\subsubsection{Mean molecular weights}

The number density of electrons is $n_e$, of hydrogen nuclei is $n_{H}$, and of 
total number of ions is $n_{ion}$. Then the total mass density is given by
$\rho=\mu_e n_e m_{H}= \mu_{H} n_{H} m_{H}= \mu_{ion} n_{ion} m_{H}=\mu_{tot} n_{tot} m_{H}$.
Here $\mu_e$ is the mean mass per electron, $\mu_{H}$ is the mean mass per hydrogen nucleus,
$\mu_{ion}$ is the mean mass per ion and $\mu_{tot}$ is the mean mass per particle (with 
 $n_{tot}=n_e+n_{ion}$ is the total number of particles).
 
\subsubsection{Characteristic time and radius scales}

 As discussed in TM99, non-radiative supernova remnants
undergo a unified evolution. The characteristic radius and time, for SNR in a uniform ISM, are given by
$R_{ch}=(M_{ej}/\rho_0)^{1/3}$ and $t_{ch}=E_0^{-1/2}M_{ej}^{5/6}\rho_0^{-1/3}$ 
with $M_{ej}$ the ejected mass and $E_0$ the explosion energy.
$\rho_0$ is the mass density of the ISM, $\rho_0=\mu_{H} n_{0} m_{H}$, where the ambient
ISM hydrogen number density is $n_0$
This gives characteristic velocity $V_{ch}=R_{ch}/t_{ch}$ and characteristic
shock temperature $T_{ch}=\frac{3}{16}\mu_I\frac{m_H}{k_B}V_{ch}^2$.
For SNR in a circumstellar medium with $\rho \propto r^{-s}$, s=2, 
the  characteristic radius and time are given by
$R_{ch}=(M_{ej}/\rho_s)$ and $t_{ch}=E_0^{-1/2}M_{ej}^{3/2}v_w/\dot{M}$,
with $\dot{M}$ and $v_w$ the wind mass loss rate and velocity, 
and $\rho_s=\frac{\dot{M}}{4\pi v_w}$.

Radiative losses gradually become important (see CMB88 for details). 
The first parcel of SNR interior gas to cool completely defines the time of thin shell 
formation $t_{sf}$. Somewhat prior to $t_{sf}$, at $t_{PDS}$, cooling causes the postshock fluid velocity
to approach the shock velocity and the pressure-driven-snowplow (PDS) phase begins. 
We use the standard cooling function $\Lambda=1.6\times10^{-19} \zeta_m T^{-1/2}$ 
erg cm$^{3}$ s$^{-1}$ with $\zeta_m$ the metallicity factor. 
This gives the same $t_{sf}$ as in CMB88, i.e. 
$t_{sf}=3.61\times10^4 E_{51}^{3/14}/(\zeta_m^{5/14}n_0^{4/7})$ yr. Following CMB88,
we define $t_{PDS}=t_{sf}/e$.

A SNR is considered to merge with the ISM when the shock speed is not significantly larger
than the ambient thermal sound speed or larger than typical turbulent velocities in the ISM.
The thermal sound speed is $c_{th}=\sqrt{\gamma k_{B} T_{ISM}/(\mu_H m_H)}$ with $\gamma$ the adiabatic index, 
here taken as 5/3, and $T_{ISM}$ the ambient ISM temperature.  The turbulent 
velocity dispersion is taken as $\sigma_{v}$. Thus merger is defined when the SNR shock speed
drops to $\beta \times c_{net}$ with  $c_{net}^2=c_{th}^2+\sigma_{v}^2$ and $\beta$ is a parameter
which determines the limit for distinguishing the SNR shock from random speeds and is taken 
to be 2 here. 
In most cases a SNR will merge before it enters the momentum conserving shell stage (CMB88). 
In this case the timescale for merger is 
$t_{mrg}=153 t_{PDS} [E_{51}^{1/14}n_0^{1/7}\zeta_m^{3/14}/(\beta c_{6})]^{10/7}$ 
with $c_{6}=c_{net}/(10^6 cm/s)$.

\subsubsection{Electron-ion temperature equilibration}
For the case of a gas consisting of a single species of particle with mass $m$, 
the post-shock temperature is $T_s=\frac{3m}{16k_B}V_s^2$.
The post shock gas is a mixture of electrons and different species of ions.
The shock wave heats the ions to a much higher temperature than the electrons
because of the high ratio of ion to electron mass ($\sim1800$ for protons, 
higher for other ions). 
We follow the discussion of electron heating given in \cite{1982CA}, which 
uses the results of \cite{1978I}. 
The electron-ion temperature equilibration timescale due to Coulomb collisions is:
$t_{eq}=5000 E_{51}^{3/14}n_0^{-4/7}$yr. 
The electron-to-ion temperature ratio is given by $g=T_e/T_{ion}$, with $g$ given to a good 
approximation by 
\begin{equation}
g=1-0.97exp[-(5f/3)^{0.4}(1+0.3(5f/3)^{0.6}].
\end{equation} 
Here $f=\frac{ln(\Lambda)}{81}\frac{4n_0}{T_s^{3/2}}(t-t_0)$ with the Coulomb logarithm
given by $ln(\Lambda)=ln(1.2\times10^5 T_s^{1/2}T_e(4n_0)^{-1/2})$, $t_0$ is the time
a parcel of gas was shocked and the post-shock density is $4n_0$.
We approximate $t-t_0$ by $t/4$ here because the mean age of the shocked gas
is $\sim t/4$.
Observations show that young supernova remnants have a lower limit on $T_e/T_{ion}$
of about 0.03 (see Fig. 2 of \cite{2013Ghav}).
In the formula for $g$ above we inserted the factor of 0.97, to obtain an early time limit of 0.03.

We apply the temperature ratio $g$ to find both electron and ion temperatures from the shock velocity. 
For a single particle species one has $T_s/m=\frac{3}{16k_B}V_s^2$.
For a mixture of species, electrons and ions, with mean mass per particle
$\mu_{e}$ and $\mu_{ion}$, respectively, one has
\begin{equation}
\frac{T_e}{\mu_e m_H}+\frac{T_{ion}}{\mu_{ion} m_H}=\frac{3}{16k_B}V_s^2.
\end{equation}
This reduces to the correct result in the case of a single species. It also allows
definition of the shock temperature, which is the temperature to which a 
(fictitious) species of mass $\mu_{tot}m_H$ would be heated, with
$1/\mu_{tot}=1/\mu_{e}+1/\mu_{ion}$.
With the electron to ion temperature ratio, $g$, and shock velocity $V_s$, one uses the 
above formulae to find $T_e$ and $T_{ion}$.

\subsubsection{Emission Measure and EM weighted temperature}
The radiation from the hot shocked plasma in the interior of a SNR is dominated by two body processes, so it
is proportional to the product of the electron and ion densities. 
(e.g. \cite{1976Ray}).
The emission measure, EM, is defined by the integral over volume of the
product of electron density $n_e$ and proton (or hydrogen ion) density $n_H$:
$EM=\int n_e(r) n_H(r) dV$.
EM depends on $n_e$, $n_H$ and their dependence on radius in the interior of the SNR, 
and on the total volume of the interior, which depends on the shock radius, $R_s$.
EM can be determined by analysis of X-ray spectrum observations of a SNR, so serves as a critical 
constraint on the state of the SNR. EM is given by the $norm$ parameter from fitting the X-ray spectrum
(see the XSPEC manual at heasarc.gsfc.nasa.gov): 
$norm=\frac{10^{-14}}{4\pi D^2}EM$, with $D$, the distance to the SNR, and all units are assumed to be cgs.
An estimate of the magnitude of EM can be done using a uniform SNR interior with typical parameters $n_e=n_H=1$ cm$^{-3}$,  $R_s=5$pc to obtain
$7.7\times10^{58}$cm$^{-3}$. Thus EM is often expressed in units of  $10^{58}$cm$^{-3}$.

To calculate EM from a SNR model, the interior density profile of the SNR is required. During the self-similar phases of
evolution of a SNR, the shape of the profile is constant with only the overall normalization and scaling with radius
changing with time. Thus we define the dimensionless EM, $dEM$, by
\begin{equation}
dEM=EM/(n_{e,s}n_{H,s}R_s^3)
\end{equation}
with $n_{e,s}$ and $n_{H,s}$ the values 
of $n_e$ and  $n_H$ immediately inside the shock front.
$dEM$ is independent of time for the self similar phases of evolution. 
Here the shock is assumed to be a strong shock with compression ratio of 4, 
so that $n_{H,s}$=$4n_H$ with  $n_H$ the ISM value.  $n_{e,s}$ is 
determined by $n_{H,s}$ and the mean molecular weights of the post-shock gas.

The observed temperature of a SNR, derived from the X-ray spectrum, 
depends on the state of the SNR and on the X-ray spectrum model used. 
For most SNRs, a single electron-temperature non-equilibrium ionization 
model is used. In this case, the observed
temperature measures the emission weighted temperature of the emitting gas.
We define the dimensionless temperature $dT$ by:
\begin{equation}
dT=\frac{1}{T_s}\frac{1}{EM}\int  n_e(r) n_H(r) T(r) dV
\end{equation}
i.e., the emission weighted temperature is $dT\times T_s$.
For the self-similar phases of evolution of an SNR, the temperature structure has a constant shape, normalized by
the temperature at the shock front, so that $dT$ is a constant. 

\subsubsection{Surface brightness profile and luminosity}

The surface brightness of a SNR depends on the line-of-sight integral of the emission coefficient, given by:
$I(b,\nu)=\int  j(\nu) ds$, where the integral is taken along the line of sight through the SNR at impact parameter $b$ from the SNR center. 
The volume emission coefficient $j(\nu)$
is given in terms of the emissivity $\epsilon(\nu)$ by  $j(\nu)=  n_e n_H \epsilon(\nu)$. 
In general $\epsilon(\nu)$ depends on the history of the particular parcel of 
emitting gas, but usually this is simplified. 
For example $\epsilon(\nu)$ can be taken as depending only on 
current temperature, $T(r)$ of the parcel. 
For simplicity, we use this approximation here.
A better approximation is that it depends on temperature and 
on an ionization timescale parameter $n_H t$ 
or, equivalently, on an ionization parameter $\eta=n_H^2 E$ (\cite{1983Ham}, \cite{2001Bork}). 

For simplicity, here we calculate the surface brightness profile using 
thermal bremsstrahlung as the emission mechanism. 
This is only a rough approximation of the
emission from an ionized plasma (e.g. see \cite{1976Ray} and \cite{1983Ham}).
The bremsstrahlung emission coefficient, in units of erg s$^{-1}$cm$^{-3}$
Hz$^{-1}$sr$^{-1}$, is 
\begin{equation}
j_B(x,dE)=5.4\times10^{-39}(Z^2)_{av} n_e(x) n_{ion}(x) T_s^{-1/2}
T_d(x)^{-1/2} ga(x,dE)exp(-dE/T_d(x)).
\end{equation}
Here $x=r/R_s$ is radius scaled by the shock radius, 
$T_d(x)=T(x)/T_s$ is interior temperature scaled by the shock temperature
and  $dE=h\nu/(k_B T_s)$ is the photon energy divided by $k_B T_s$.
Thus the argument of the exponential reduces to $h\nu/(k_B T(x))$. 
$ga$ is the Gaunt factor and $(Z^2)_{av}$ is the mean value of $Z^2$ for the plasma.
$n_e(x)$ and $n_{ion}(x)$ are given by the interior density profile of 
the SNR, normalized to 1 at the shock front, times the values at the 
shock front $n_{e,s}$ or $n_{ion,s}$.
Thus we obtain the surface brightness at a given scaled photon energy $dE$:
\begin{equation}
I(b,dE)=\int  j_B(x(s),dE) ds
\end{equation}
where $x(s)=\sqrt{b^2+s^2}/R_s$ is the 
scaled radius along the line of sight at a given impact parameter $b$.
The total luminosity of the SNR at a given scaled photon energy $dE$ is 
given by integrating the surface brightness over the area of the SNR, which 
is equivalent to integrating the emission coefficient over volume of the SNR: 
$L_{\nu}(dE)=\int_{0}^{R_s} I(b,dE) 2\pi b db$.
The luminosity in any given energy band $E_1$ to $E_2$ is obtained by 
integrating $L_{\nu}$ over the scaled energy range $dE_1=E_1/(k_B T_s)$
to $dE_2=E_2/(k_B T_s)$.

\subsection{Standard evolutionary path}

\subsubsection{Ejecta Dominated (ED) and Sedov-Taylor (ST) stages}

The ED phase is the time during which the ejected mass dominates over the swept-up mass from
the CSM on the evolution of the SNR. 
The ST phase is the time after the effects of ejected mass are dominant but prior to onset of
radiative losses. We refer to the so-called standard ST solution, which assumes zero ejected
mass, as "pure ST" to differentiate it from the ST phases calculated by TM99, which include
the effects of non-zero ejected mass.

The evolution starts with the early ED self-similar phase, previously
considered by \cite{1982Chev} for $n>5$. 
The shock evolution during the full ED and ST phases can be calculated using the results of TM99. 
They consider the full non-radiative evolution of a SNR, which ends with the start of
 transition to the PDS radiative phase at time $t_{pds}$. 
During the ED phase, the reverse shock propagates inward into the ejecta, eventually
reaching the center of the SNR. A core-envelope structure for the ejecta is assumed
with constant density core and power-law density envelope, with index n. 
For $n<3$, the core can be taken to have zero size. For $n>5$ the outermost velocity of the
envelope $v_{ej}$ does not affect the mass or energy of the envelope so that the parameter
$v_{core}/v_{ej}$ can be taken to be zero (by taking the limit $v_{ej}=\infty$).
Here we calculate SNR forward and reverse shock evolution for uniform CSM (s=0) for the following
cases of ejecta power law index: n=0, 2, 4, 6, 7, 8, 9, 10, 12 and 14. 
For the CSM wind profile case(s=2), we calculate n=0, 1, 2 and 7 cases.

First consider the s=0 cases. For n=0 we use the solutions given in Table 5 of TM99.
For this case the reverse shock reaches maximum radius at 1.046$t_{ch}$. 
The early self-similar evolution, with constant ratio of reverse shock radius $R_r$ to 
forward shock radius $R_b$, ends much earlier: it is violated at the 5\% level
by 0.175$t_{ch}$.
For n=2, we use the solutions of TM99 with parameters given in their Table 3, noting that
their Table 4 solutions contain some incorrect numbers, inconsistent with Table 3.
For $t<t_{ST}$, the solutions are specified by $t(R_b)$ and $t(R_r)$, so these are inverted
using a root finding algorithm to yield $R_b(t)$ and $R_r(t)$, hence $V_b(t)$ and $V_r(t)$.
For n=4 and $t<t_{ST}$, we use the TM99 equations (47), (48) and (49).
For n=4 and $t>t_{ST}$, we use the TM99 equations (57) and (58).
However to better match their n=4 numerical solutions for the reverse shock given in their 
Figs. 4 and 5, we add an acceleration term to $R_r(t)$ for $t>t_{core}$.
As for n=2, for $t<t_{ST}$ the relations $t(R_b)$ and $t(R_r)$ are inverted to get the solutions.
For n=6, 7, 8, 9, 10, 12 and 14, we use the $n>5$ solutions of TM99. 
For $R_b(t)$, the solution is specified by their equations (75) and (76)  for $t<t_{ST}$
and equations (57), (81) and (82) for $t>t_{ST}$.
For $R_r(t)$, the solution is specified by their equation (77) and $R_r=R_b/l_{ED}$ for $t<t_{core}$
and equations (83) and (84) for $t>t_{core}$ until the reverse shock hits the center.

Next we discuss the s=2 cases. For n= 0, 1 and 2 we use the solutions from Table 8 of TM99
and invert $t(R_b)$ to get $R_b(t)$ then use  $R_r=R_b/l_{ED}$.
For n= 7 we use the solutions from Table 9 of TM99, which is the same as that from \cite{1982Chev}.

\subsubsection{Pressure-driven snowplow (PDS) phase, momentum-conserving shell (MCS) phase and merger}

We follow the treatment given in CMB88 for the radiative phases for SNR in a uniform
CSM (s=0 case). 
The different cases above with different ejecta profiles (values of n) have radii
different at a given time than the pure Sedov-Taylor solution (see Results section).
Thus the matching of radiative phases onto the non-radiative phase has to be done differently
than in CMB88. 
CMB88 investigated numerical hydrodynamic solutions of SNR evolution, then
compared those to analytic models for blast wave evolution.
They then fit an offset power-law as a good approximation to the numerical solutions. 
However this creates a discontinuity in velocity at $t_{pds}$ where the relation 
$V_b=dR_b/dt$ is violated.
In order to get a smooth evolution of the blast wave radius and velocity and also satisfy
$V_b=dR_b/dt$ at all times, we include a smooth transition.
We linearly interpolate the shock velocity from the value at $t_{pds}$ from the previous 
phase to the value in the PDS phase at time 1.1$t_{pds}$.
Then we integrate the equation $V_b=dR_b/dt$ to get $R_b(t)$ at all times for $t>t_{pds}$.

To determine when the PDS phase ends we calculate the merger time $t_{mrg}$ and the onset
of the MCS phase $t_{mcs}$, as defined in CMB88. 
If $t_{mrg}<t_{mcs}$ then the evolution is terminated.
If $t_{mrg}>t_{mcs}$ then $V_b$ for the MCS phase is calculated similarly to that for 
the beginning of the PDS phase, by interpolating between $t_{mcs}$ and 1.1$t_{mcs}$ and
then integrating $V_b=dR_b/dt$. 

\subsection{Alternate SNR evolution models}

\subsubsection{SNR in cloudy ISM}
We follow the treatment of WL91, which presented self-similar models for SNR
evolution in a cloudy ISM.
The self-similar models are similar to the standard self-similar Sedov-Taylor model for
the evolution of the blast wave radius, although the interior structure of the
cloudy SNR models is drastically different.
The WL models depend on two parameters: a cloud density parameter $C=\rho_c/\rho_0$, 
with $\rho_c$ is the ISM density if the clouds were uniformly dispersed in the ISM; 
and an evaporation timescale parameter $\tau=t_{evap}/t$, with $t_{evap}$ the evaporation
timescale and $t$ the age of the SNR.
The WL models reduce to a one parameter set which depend on $C/\tau$, 
in the limit $\tau$ approaches $\infty$. We use these one parameter models, in
particular, the three cases $C/\tau$=1, 2 and 4. 
The WL91 model is strictly applicable, like the standard Sedov-Taylor model, for the case
of zero ejected mass, and evolution in a cloudy ISM from t=0. 
We join the cloudy SNR models onto the earlier phases with finite ejected mass  at time $t_{ST}$. 
Earlier than $t_{ST}$, the SNR blast wave evolution is dominated by the interaction of the
ejecta, so the effects of a cloudy ISM are hard to determine and have not yet 
been investigated in enough detail.
The transition to cloudy SNR evolution is calculated by interpolating the forward
shock velocity $V_b$ at $t_{ST}$, to the velocity given by the WL91 model at 1.1$t_{ST}$.
Then $R_b$ is found by integrating $V_b=dR_b/dt$ for $t>t_{ST}$.

The cloudy SNR phase ends when radiative cooling becomes important. However it ends
very differently than for the ST phases described above.
Rather than entering a PDS phase, the SNR enters directly into the MCS phase. 
This happens because of the interior structure for the WL91 models  (see Fig.2 of WL91).
Because the temperature and density are much more uniform than for the ST
models (pure ST model and ST models of TM99), the whole interior of the cloudy SNR cools at nearly the same time, which
results in the loss of interior pressure at nearly the same time that the forward
shock evolution is affected by loss of pressure.

\subsubsection{Fractional energy loss model}

 LK2000 introduced an analytic model for radiative blast wave evolution, based
on the thin shell approximation. They introduced a parameter $\gamma_1$ which determines
the fractional energy loss rate at the shock front, $\epsilon$ by the relation:
$\epsilon=4\frac{\gamma-\gamma_1}{\gamma-1}(\gamma_1+1)^{-2}$. Here
$\gamma$ is the adiabatic index of the gas (both upstream and downstream) and
$\gamma_1$ is constrained by $1\le \gamma_1 \le \gamma$.
The post-shock expressions for pressure, velocity and density are the conventional
expressions but with $\gamma$ replaced by $\gamma_1$ (see LK2000).
The total energy content of the blast wave decreases with time as a result of
the energy loss: $dE/dt=-2\pi\rho_0 V_b^3 R_b^2 \epsilon$.
The momentum conservation equation, with internal driving pressure, determines the motion
of the blast wave: $dV_b/dR_b=3(\alpha-1)V_bR_b$,
with the ratio of interior pressure $p_c$ to post-shock pressure $p_f$ given by
$\alpha=p_c/p_f$.
The total energy of the SNR is given by the kinetic energy of the shell plus the internal energy of the hot interior (equation 6 of LK2000). 
Then $\alpha$ is given by $\alpha= (2-\gamma+[(2-\gamma)^2+4(\gamma_1 -1)]^{1/2})/4$.
As in LK2000 we define $n_{\alpha}=1/(4-3\alpha)$, then integrate the equation of motion
for $V_b$. 
However we use different boundary conditions than LK2000 so obtain different solutions for
$R_b(t)$ and $V_b(t)$. 
I.e. we join the constant energy loss solution onto an earlier phase of
SNR evolution at time $t_0$, with radius $R_0$ and velocity $V_0$. 
The resulting solution is:
\begin{equation}
R_b(t)=[R_0^{1/n_{\alpha}}+(4-3\alpha)R_0^{3-3\alpha}V_0(t-t_0)]^{n_{\alpha}}
\end{equation}
and 
\begin{equation}
V_b(t)=V_0[R_b(t)/R_0]^{3\alpha-3}.
\end{equation}

The above solution was derived using the momentum equation for a thin shell.
However, when one sets $\epsilon=0$ and uses initial $t_0$, $R_0$ and $V_0$ from 
the pure ST stage, one recovers the pure ST evolution for $R_b(t)$ and $V_b(t)$.
The evolution of $R_b(t)$ and $V_b(t)$ for the constant energy loss model with $\epsilon=0$ also reduces to
the TM99 evolution for the various n values in the ST phase.
When choosing a non-zero fractional energy loss $\epsilon>0$, the shock front decelerates
more rapidly than for the respective adiabatic phases (pure ST phase or TM99 ST phases).
This is expected because of the reduced post-shock pressure resulting from the energy loss.
In our SNR modelling software, we allow the user to have
fractional energy loss phase start at any time in the ST or PDS phases.
Prior to $t_{ST}$, the fractional energy loss model does not properly represent the SNR evolution.
 
\subsubsection{SNR in hot low-density medium}

The evolution of a SNR in a hot low-density medium deviates significantly from the 
ST evolution because of the effects of swept-up energy from 
the ISM and of the high sound speed \cite{2005TW}(hereafter TW2005).
They verified their analytic approximations with hydrodynamic simulations.
The effects of a hot ISM are large for the evolution if the critical time $t_c$
is less than the shell formation timescale for the SNR, $t_{sf}$.
$t_c$ is defined by $t_c=[(2\zeta_1/5)^5 E_0/(\rho_0 c_{th}^5)]^{1/3}$ 
with $\zeta_1=2.026^{1/5}$ (see TW2005).
If $t_c<t_{sf}$, then the SNR never enters the radiative phase but instead has an outer
velocity which asymptotically approaches the sound speed in the hot ISM.

Instead of matching the hot-ISM SNR solution onto a pure ST model (as done in TW2005),
we match to any of the TM99 ST phases at time $t_M$, which has a default value of $0.1t_c$.
At earlier times than $\simeq 0.1t_c$, the analytic solution is not a good match to the 
hydrodynamic solutions (see TW2005 Fig.1).
This gives the following evolution for SNR outer velocity:
\begin{equation}
V_H(t)=c_{th}[t_H/t+1]^{3/5} 
\end{equation}
with $t_H=[(V_b(t_M)/c_{th})^{5/3}-1]t_M$.
The SNR outer radius is found by integration of the velocity:
\begin{equation}
R_H(t)=R_M + \int_{t_M}^{t} V_H(t) dt.
\end{equation}

\subsection{Implementation notes}

The models were originally programmed in MathCad and then programmed in
Python. For debugging and verification of the calculations, the results 
from both codes were compared for a extensive of cases. 

The input parameters for the SNR model calculations are as follows.
The SNR explosion parameters are: age (in yr),
energy $E_0$ (units $10^{51}$erg), ejected mass $M_{ej}$ (units $M_{\odot}$),
ejecta power-law index n. 
The ambient medium parameters are:
power-law index s (0 or 2), temperature $T_{ISM}$, 
cooling adjustment factor $\zeta_m$ (1 for solar abundance) and
ISM turbulence or random speeds $\sigma_v$ (km/s).
For s=0, density is specified by ISM number density.
For s=2, input parameters are wind mass loss rate ($M_{\odot}$/yr) and wind speed (km/s).  
The electron to ion temperature ratio is calculated as described above, 
but alternately can be entered by hand.  

Element abundances for the ISM or CSM and for the ejecta are specified by user input.
For simplicity, in the program we include the elements H, He, C, N, O, Ne, Mg, Si, S and Fe, which are the ten most abundant elements in Solar abundances. 
The abundances are specified in standard form, i.e. by $log(X/H)+12$ 
with $X/H$ is the number of nuclei of element $X$ divided by the number of $H$
nuclei.
The default values for the ISM abundances are solar values, taken from \cite{1998GS},
or LMC values, taken from \cite{RD1992}.
For ejecta abundances, default values were taken (in units of $log(X/H)+12$) as follows:
for core-collapse SN-
 He: 11.22, C: 9.25, N: 8.62, O: 9.69, Ne: 8.92, Mg: 8.30, Si: 8.79,
 S: 8.54, Fe: 8.55; 
for Type Ia SN-
 He: 11.40, C: 12.60, N: 7.50, O: 12.91, Ne: 11.04, Mg: 11.55, Si: 12.75,
 S: 12.43, Fe: 13.12.
The pop-up window that specifies abundances allows individual abundances for
the 9 elements relative to $H$ to be specified by the user.  

Additional inputs for the alternate models, which all require s=0, are as follows.
For the cloudy ISM model, the input is  $C/\tau$ (1, 2, or 4).
For the fractional energy loss model, inputs are adiabatic index $\gamma$, 
fractional energy loss $\epsilon$, and model start time (within ST or PDS phases).
For the hot low-density ISM model (available if $0.1t_c<t_{pds}$) the input is the end time.

Outputs from the program are as follows.
Plots are produced for $R_b$ and $R_r$ (if applicable) vs. time, and
 for $V_b$ and $V_r$ (if applicable) vs. time.
 Values are given, at the specified age, for blast-wave shock temperature for electrons, 
 reverse shock temperature for electrons, blast-wave radius and velocity, reverse shock radius and velocity.
 Also given are the transition times between ED and ST phases, between ST and PDS phases,
 between PDS and MCS phases (if applicable), and merger time.
 Emissivity related values and luminosity are calculated only for phases
 for which the internal structure (density and temperature) are determined
 by self-similar models. 
 Those phases are: the ED phase for $t<t_{core}$ (see TM99);
 the ST phase for $t>t_{rev}$ where $t_{rev}$ is the time that the reverse shock
 hits the center of the SNR; and the cloudy ISM models (see WL91).
 As noted in Section 3.1.4 above, for the self-similar phases the dimensionless
 emission measure and dimensionless temperature are constant with time.
 These values (dEM and dT) were calculated using the self-similar interior
 solutions for density and temperature and given in Table 1 below.
 For the ED phase, the values for the forward and reverse shocked material
 are given for the two cases s=0, n=7 and s=0, n=12.
 
\section{Results and Discussion}

The model calculations were tested by verifying  results in TM99, CMB88, WL91, LK2000 and TW2005.
The MathCad version of the program and Python version were compared for verification. 
The MathCad version was recently used to model the set of 50 SNRs for which 
good observations were available, and to derive properties of this set of
SNRs (\cite{2016Leahy}).
To illustrate the SNR evolution modelling software, we present the following cases.

\begin{figure}
\plotone{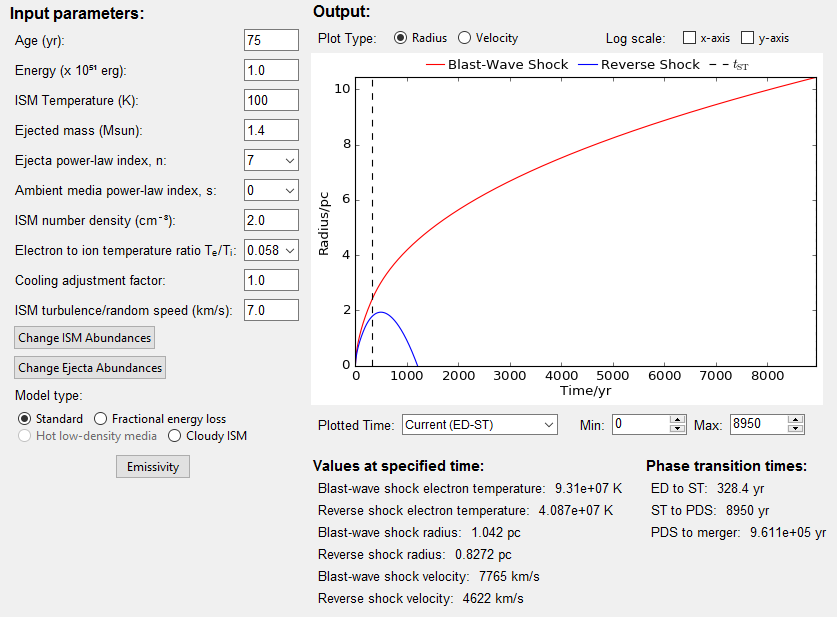}
\caption{Screenshot of the main screen of the SNR modelling program. Left hand side: 
input parameters (boxes). Buttons specify the model type or bring up
windows for entering ISM and ejecta abundances.
Upper right: output plot of radius or velocity vs. time. 
Lower right: output values at the specified time and the phase transition times. 
Lower left: the Emissivity button brings up the calculation of emissivity and luminosity.
\label{figMain}}
\end{figure}

The main window of the program takes input parameters required to specify 
the SNR evolution, see Fig. \ref{figMain}. 
ISM and ejecta abundances can be selected in a pop-up menu from 
standard abundance sets or else the element abundances can be specified individually.  
The model type can be selected from the models described above: 
either the standard SNR evolution,
fractional energy loss model, cloudy ISM model or Hot low-density ISM model.
The main window produces several outputs. These include a plot
of shock radius or velocity vs. time; shock radius, velocity and electron temperature for the specified input age; and the phase transition times. 

\begin{figure}
\plottwo{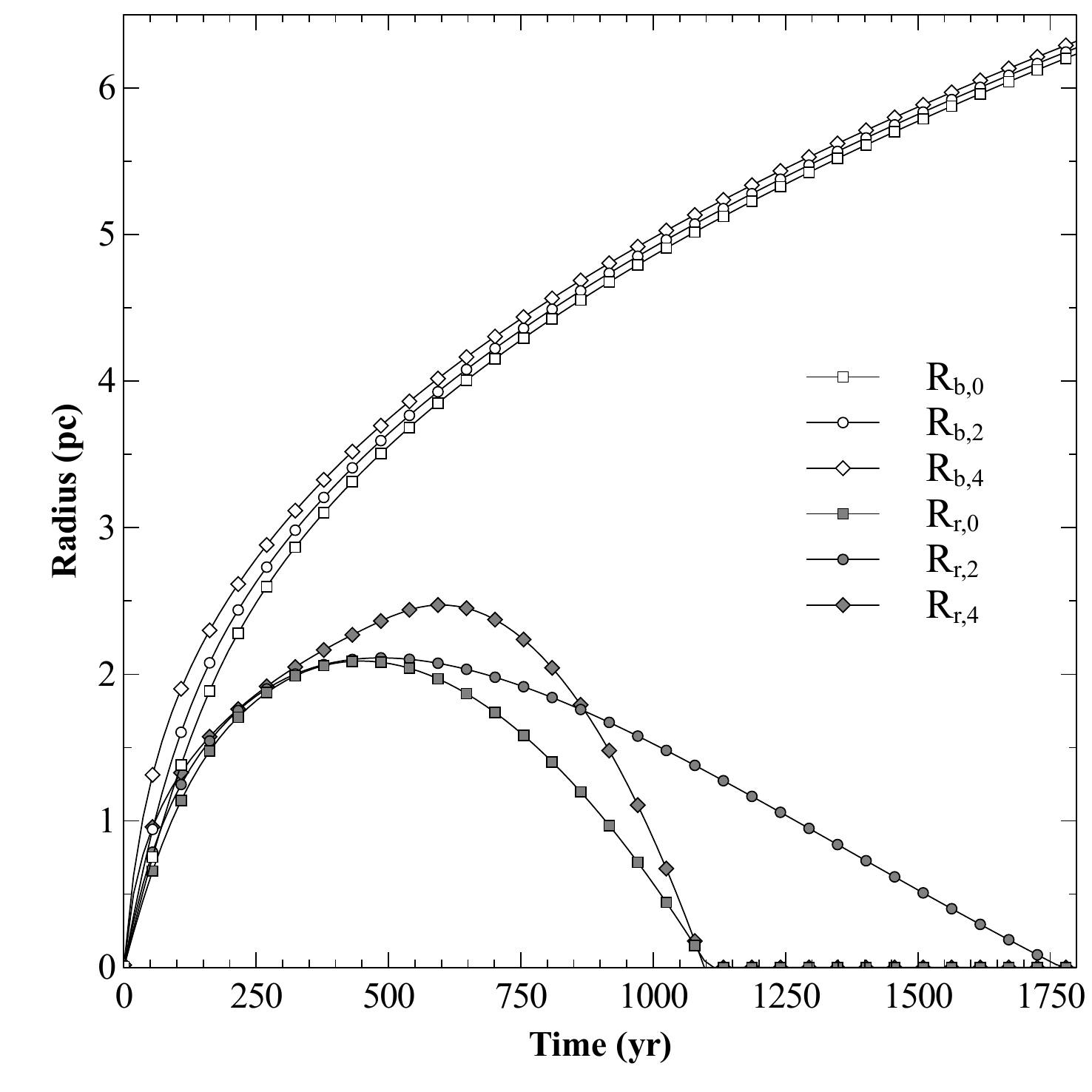}{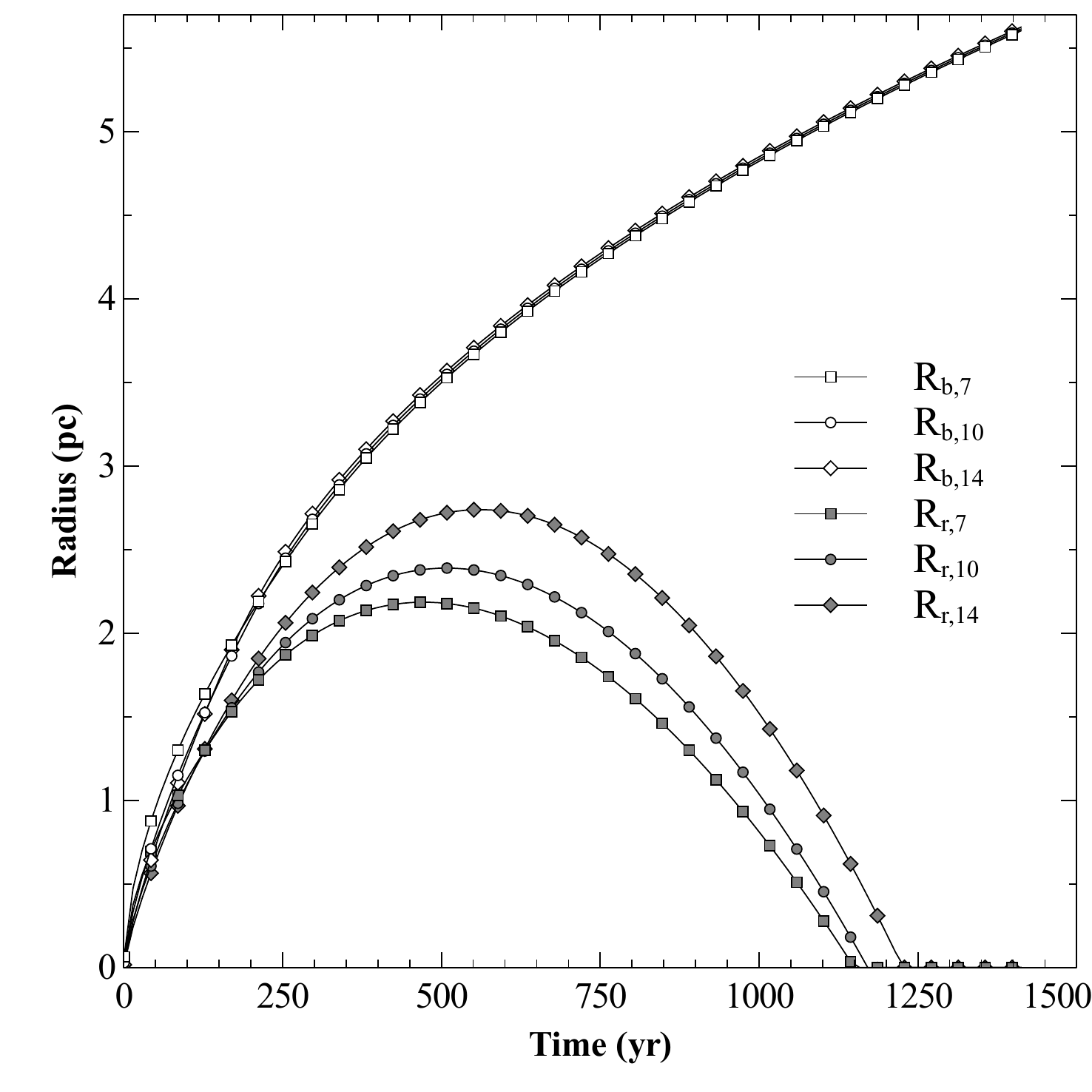}
\caption{
The time evolution of blast-wave radius ($R_b$) and reverse shock radius ($R_r$)
vs. time: for ejecta density profiles with n= 0, 2 and 4 (left panel) and for
 for ejecta density profiles with n= 7, 10 and 14 (right panel). 
 \label{figED}}
\end{figure}

For the ED and early ST phases, we compare calculations of forward (or blastwave) and reverse shock radii, $R_b$ and $R_r$, in Fig. \ref{figED}. 
Cases with s=0 and n=0, 2, 4, 7, 10 or 14 are shown. 
Input parameters of $E_0=10^{51}$ erg, $M_{ej}=1.0~M_{\odot}$, $n_0=2$ cm$^{-3}$,
$T_{ISM}=100$ K, $\zeta_m=1$ and $\sigma_v=7$ km/s were used.
The forward shock radius is largest for n=4 and smallest for n=7 (see also
Fig. \ref{figSTvSed} below). 
The reverse shock lifetime is longest for the n=2 case, but 
reaches the largest radius for the n=14 case.

\begin{figure}
\plotone{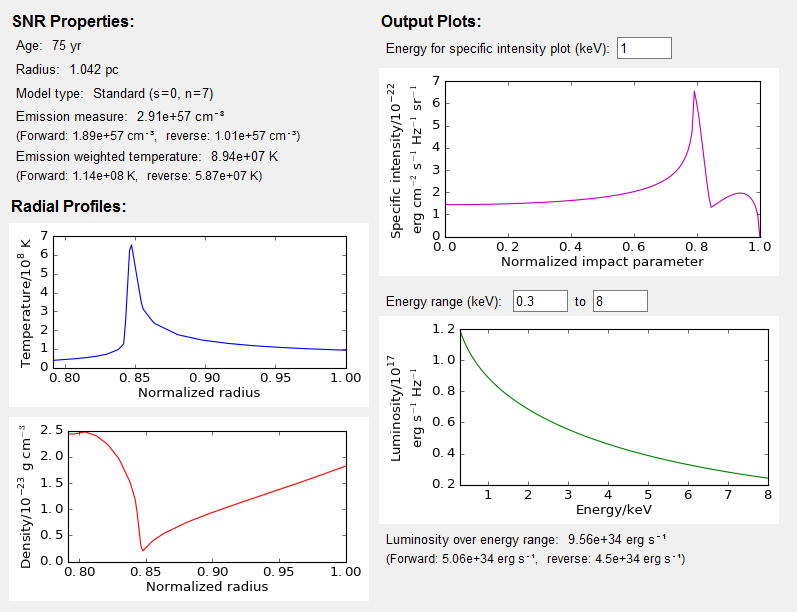}
\caption{Screenshot of the Emissivity calculation tool of the SNR modelling program:
For a SNR in the ED phase at age 75 yr with ejecta power law index n=7 in a uniform CSM (s=0): radial profile of temperature (top left) and of density (bottom left).  Surface brightness profile at 1 keV 
(top right panel) and spectrum integrated over the volume of the SNR (bottom right panel). 
\label{figEMED}}
\end{figure}

The Emissivity button on the main screen of the program brings up
calculation of various quantities calculated from the interior structure of
the SNR.
Sample results for the ED phase are shown by the screenshot in Fig. \ref{figEMED}, for the case s=0, n=7. 
We use ejecta composition typical for core collapse SN and solar abundances for the ISM.
Emission measures are given for the forward shocked material, between
the forward shock and the contact discontinuity, and for the reverse
shocked material, between
the reverse shock and the contact discontinuity. 
For this case the forward shock EM is larger by a factor of 1.9.
In general, the emission weighted temperature is 
different than the temperature at the shock front.
This is caused by the temperature and density profiles behind the
shock front.
In this case, the electron temperature at the forward shock front is 
$9.3\times10^7$K compared to the emission weighted temperature of  
$11.4\times10^7$K. 
For the reverse shock
the electron temperature at the reverse shock is $4.1\times10^7$K 
whereas the emission weighted temperature of reverse shocked material is 
$5.9\times10^7$K.
Radial profiles of temperature and density are given as a function of $r/R_s$.
The surface brightness profile $I(b,E)$, using the bremmstrahlung emissivity
as described above, is given for a user specified 
photon energy, $E$, in keV.
The SNR spectrum $L_{\nu}$ is calculated by integrating $I(b,E)$ over the 
face of the SNR for a given input energy range, and the total luminosity
$L$ over the same energy range is calculated. 
The luminosity from the forward shocked material and reverse shocked
material are calculated separately.

Next we consider the ST phase and compare the shock radii and
velocities for different ejecta profiles, n, and with the pure ST model.
We find that there is a significant difference between pure ST
and  ST evolution for different ejecta profiles n. 
The left panel of Fig. \ref{figSTvSed} shows blast wave radius, calculated with the input parameters above.
We note that pure ST is equivalent to setting the ejecta mass equal to zero.
Both the ejected mass and how it is distributed affect the evolution of  $R_b$:
the different cases of ejecta power-law index, n, differ by similar amounts to the difference with pure ST.
For n=7, the forward shock radius $R_b$ is smaller than for pure ST by  8\%, 3.7\% and
0.8\%  at $t_{ch}$, $2t_{ch}$ and $10t_{ch}$, respectively.
The \% difference decreases with time, although absolute difference is nearly constant in time.
The shock velocities $V_b$ differ by even more with differences of 13\%, 6\% and 1.1\%
for the same three times.
The outer shock radius ($R_b$) evolution for n=10 to 14 is nearly identical (see Fig. \ref{figED} right panel).
Thus these cases differ the most from pure ST (Fig. \ref{figSTvSed} left panel). If one uses a pure ST model
and uses radius as an age proxy, the SNR age will be underestimated compare to the more accurate n=7 to n=14 models.
The age error is significant, at about 15\%.
The velocities for the various cases are compared in the right panel of Fig.  \ref{figSTvSed}. 
It is seen the ST evolution for various n has larger velocity at a given time than pure ST.
If one uses velocity as an age proxy, the SNR age will be underestimated using pure ST compared to the more accurate 
n=7 to n=14 models, similar to the case for using radius as an age proxy.

An example of outputs of the Emissivity calculation for the ST phase 
is shown by the screenshot in Fig. \ref{figEMSed}, for the case s=0, n=7. 
Emission measure and emission weighted temperature are calculated.
The electron temperature at the forward shock front is 
$2.37\times10^7$K compared to the emission weighted temperature of  
$3.05\times10^7$K. 
Radial profiles of temperature and density are given as a function of $r/R_s$.
The surface brightness profile $I(b,E)$, using the bremmstrahlung emissivity, 
is given for the input specified energy, in this case 1 keV.
The SNR spectrum $L_{\nu}$ is calculated over the input energy range of 0.3 to 8 keV, and the total luminosity $L$ over the same energy range is calculated. 

The late-time SNR evolution is illustrated in Fig. \ref{figPDSMCS}. 
In this case to obtain the MCS phase to occur prior to merger, 
the parameters were taken as $E_0=10^{50}$ erg, 
$M_{ej}=1.0~M_{\odot}$, $n_0=0.1$ cm${-3}$, $T_{ISM}=30$ K, $\zeta_m=2$ 
and $\sigma_v=0.3$ km/s.
The left panel shows shock radius $R_b(t)$ for the ST phase n=0 case (labelled $R_{b,0}$),
for the PDS solution joined to ST n=0 solution (labelled $R_{pds}$),
for the full joined solution (labelled $R_{mcs}$).
The latter includes ST, PDS and MCS evolution.
The decrease of shock speed at the ST to PDS transition (at $\sim2.4\times10^4$ yr) 
and decrease at the PDS to MCS transition (at $\sim10^7$ yr) are clearly seen.
In the right panel, the deceleration parameter $m=V_bt/R_b$ is shown for the 
full joined solution (ST to PDS to MCS). 
During the ST to PDS evolution (between $10^4$ to $10^5$ yr), $m$ drops from the pure ST value
of 0.4 to $\simeq$0.3 for the PDS phase. 
At the end of the PDS phase there is a rapid drop to a value of $\simeq0.16$ followed
by a slow increase.
The behaviour of a gradual drop of $m$ from 0.4 to 0.3 for the ST to PDS transition was found
by CMB88 (their Fig. 5). 
They also note that practically (i.e. for most cases of realistic input parameters) the MCS stage does not occur.
This is in agreement with the current calculations: to obtain an MCS stage we had
to use a small explosion energy and low ISM density, and
we also had to use a cold ($T_{ISM}<40$K) and very quiet ($\sigma_v<1$ km/s) ISM.
The latter two restrictions are unrealistic for a low density ISM, confirming that the MCS phase
can occur only in rare circumstances.

We illustrate the effect of evolution of a SNR in a cloudy medium next.
In the left panel of Fig. \ref{figCloudy}, shock radii for
the pure ST model, the ST n=0 model and cloudy SNR models are compared.
The cloudy SNR models are calculated for $C/\tau=1$, 2 and 4. These cloudy
SNR models are all joined at small t onto the ST n=0 model.
The radius of the cloudy SNR is much smaller than pure ST (by 8\% for
$C/\tau=1$, up to 20\% for $C/\tau=4$, with the difference
nearly constant in time). The difference between the cloudy models and 
the ST n=0 model is similar.
In the right panel of Fig. \ref{figCloudy} the corresponding shock velocities
of the different models are compared. This shows that the shock
velocities are lower for the cloudy models (by $\sim$10\% to 20\%) 
than the pure ST model or ST n=0 model.

The evolution of a SNR occurring in a hot ISM is illustrated in Fig. \ref{fighotISM}.
The input parameters for this case are: $E_0=10^{51}$ erg, 
$M_{ej}=1.0~M_{\odot}$, $n_0=1$ cm$^{-3}$, $T_{ISM}=10^6$ K, $\zeta_m=0.1$ and $\sigma_v=7$ km/s.
The left panel of Fig. \ref{fighotISM} shows the evolution of blast-wave radius for the case of 
 the standard evolution, without including the effect of swept-up ISM thermal energy for the ST phase, $R_{b,ST}$, followed by the PDS phase, $R_{PDS}$, and ending with merger. 
Recall that time of merger with the ISM was defined as the time that shock velocity drops to  $\beta \times c_{net}$
where  net speed of sound $c_{net}$ includes thermal and random motions, 
and  $\beta$ was taken as 2.
The left panel also shows shock radius for the hot ISM model, $R_H$. 
It lasts significantly longer.
The right panel of Fig. \ref{fighotISM} shows the shock velocities for
the standard model ($V_{b,ST}$ and $V_{PDS}$) and for the hot ISM model ($V_{H}$).
The deceleration of velocity is much slower for the hot ISM model, after about 
$1.5\times10^4$ yr. 
This results in the velocity reaching the merger velocity at much later time.
For a slightly cooler ISM, $T_{ISM}=0.5\times10^6$K, the swept up energy of the ISM is not enough to 
prevent the standard evolution from taking place (PDS and merger occurs before the shock velocity is
affected significantly). 

\section{Summary and Conclusion}

The different evolutionary phases for spherically symmetric SNR evolution have been described by analytical solutions. The phases include the full evolution, 
from the early ejecta-dominated stage to the merger with the interstellar medium.
Joining of the different phases of evolution has been done in a consistent way,
so that a continuous evolution in shock radius and velocity is obtained for the
 whole SNR evolution. In addition to the standard, optional evolutionary models 
 have been described, including evolution in a cloudy ISM, the fractional energy 
loss model, and evolution in a hot low-density ISM.

These analytic descriptions have been incorporated into a Python code which has 
a graphical interface. The interface takes the SN explosion parameters and the
 CSM/ISM parameters as inputs and produces a number of outputs. The outputs 
 include plots of shock radius and velocity vs. time, transition times between
 evolutionary phases and other useful information. It also includes the SNR 
 surface brightness profile at a specified photon energy, the SNR integrated
 spectrum and luminosity.
 
 The program has been used here to illustrate some properties of SNR evolution, 
 such as the significant different between shock radius and velocity between pure
 Sedov-Taylor evolution and more realistic models which include the effects of
 ejected mass. 
 This software is intended for two purposes: 
 for understanding SNR evolution; and for modelling SNR observations to
 determine good estimates of SN explosion properties, SNR evolutionary state and
ISM properties. The program is available in two forms, as Python code or 
as a windows executable. It can be obtained from GitHub (repository denisleahy/SNRmodels) or from www.quarknova.ca (under software). The README.txt file
 has installation instructions. Those who use the software
are requested to reference the current publication in any published work.
  
\acknowledgments

This work was supported by a grant from the Natural Sciences and Engineering Research Council of Canada. 

\clearpage

\clearpage

\begin{figure}
\plottwo{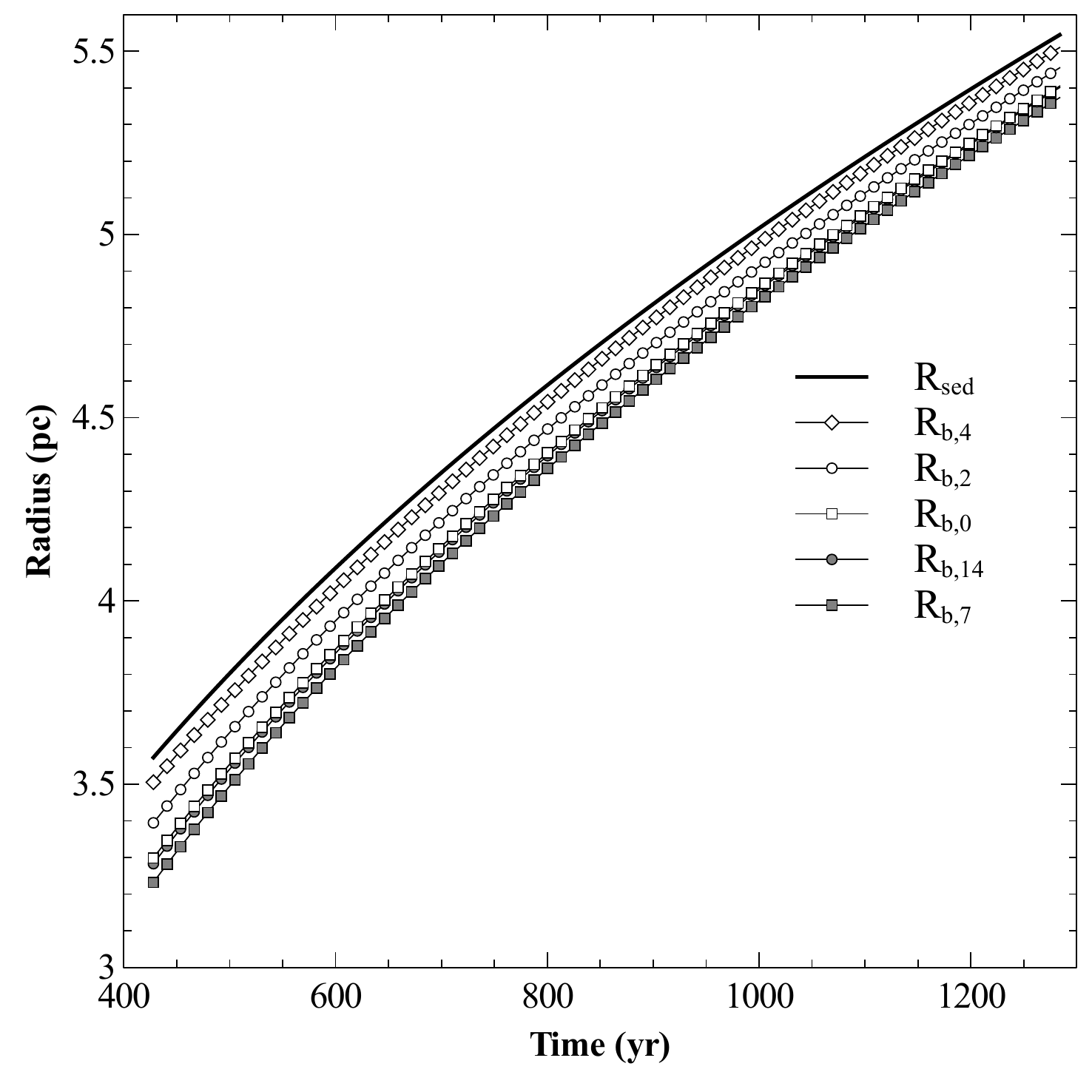}{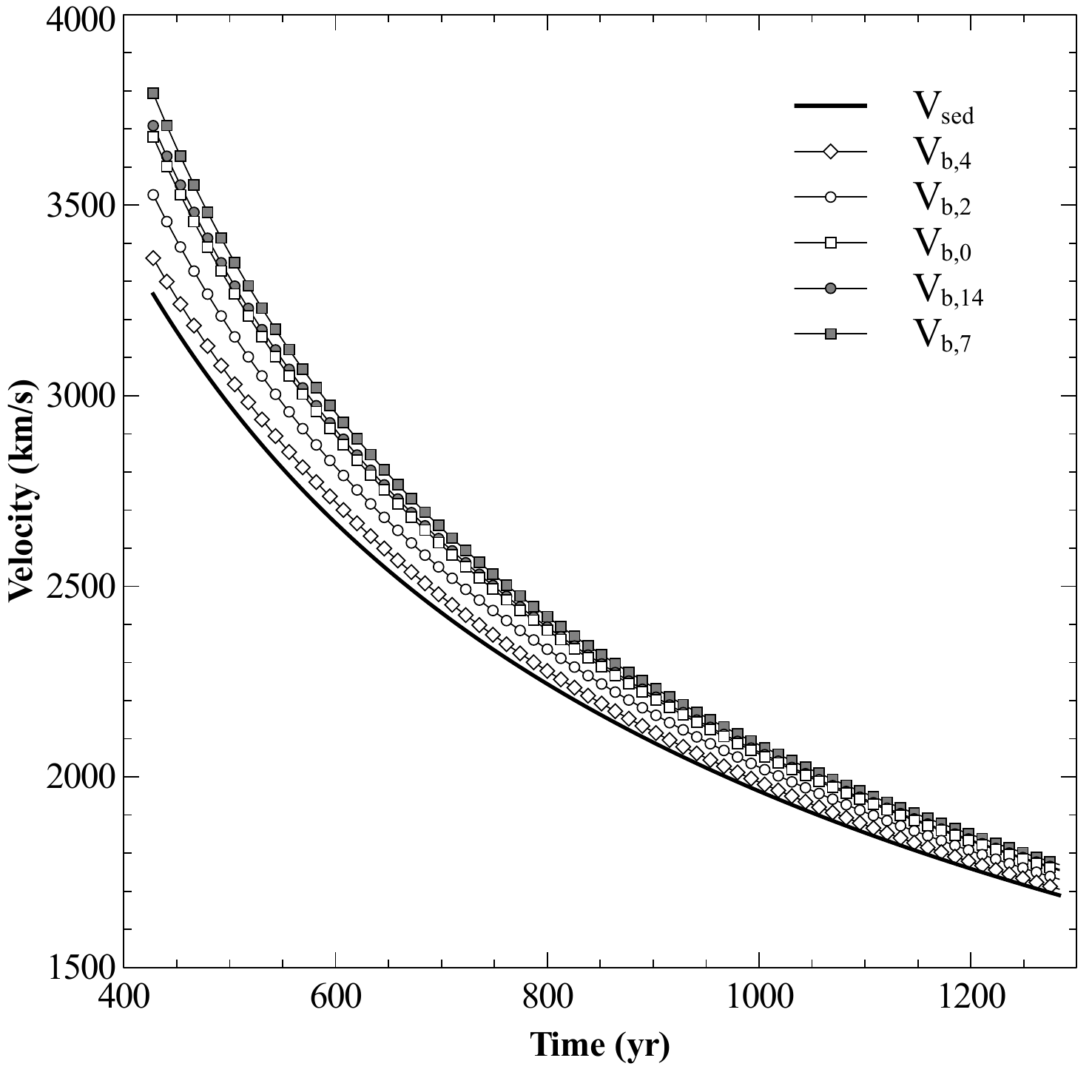}
\caption{
The time evolution of blast-wave radius ($R_b$) and velocity $V_b$
vs. time. Left panel: comparison of pure Sedov-Taylor solution $R_{sed}$ with $R_b$ for ejecta density profiles with n= 0, 2, 4, 7, 10 and 14.  
Right panel:  comparison of shock velocities: pure Sedov-Taylor solution $V_{sed}$ with $V_b$ for ejecta density profiles with n= 0, 2, 4, 7, 10 and 14.
\label{figSTvSed}}
\end{figure}

\begin{figure}
\plotone{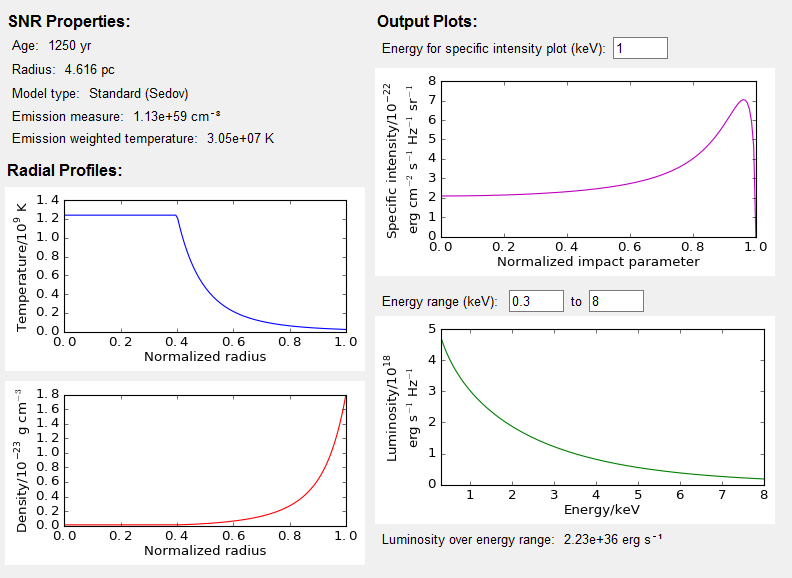}
\caption{
Screenshot of the Emissivity calculation tool of the SNR modelling program:
For a SNR in the ST phase at age 1250 yr in a uniform CSM (s=0): radial profile of temperature (top left) and of density (bottom left).  Surface brightness profile at 1 keV 
(top right panel) and spectrum integrated over the volume of the SNR (bottom right panel). 
\label{figEMSed}}
\end{figure}

\begin{figure}
\plottwo{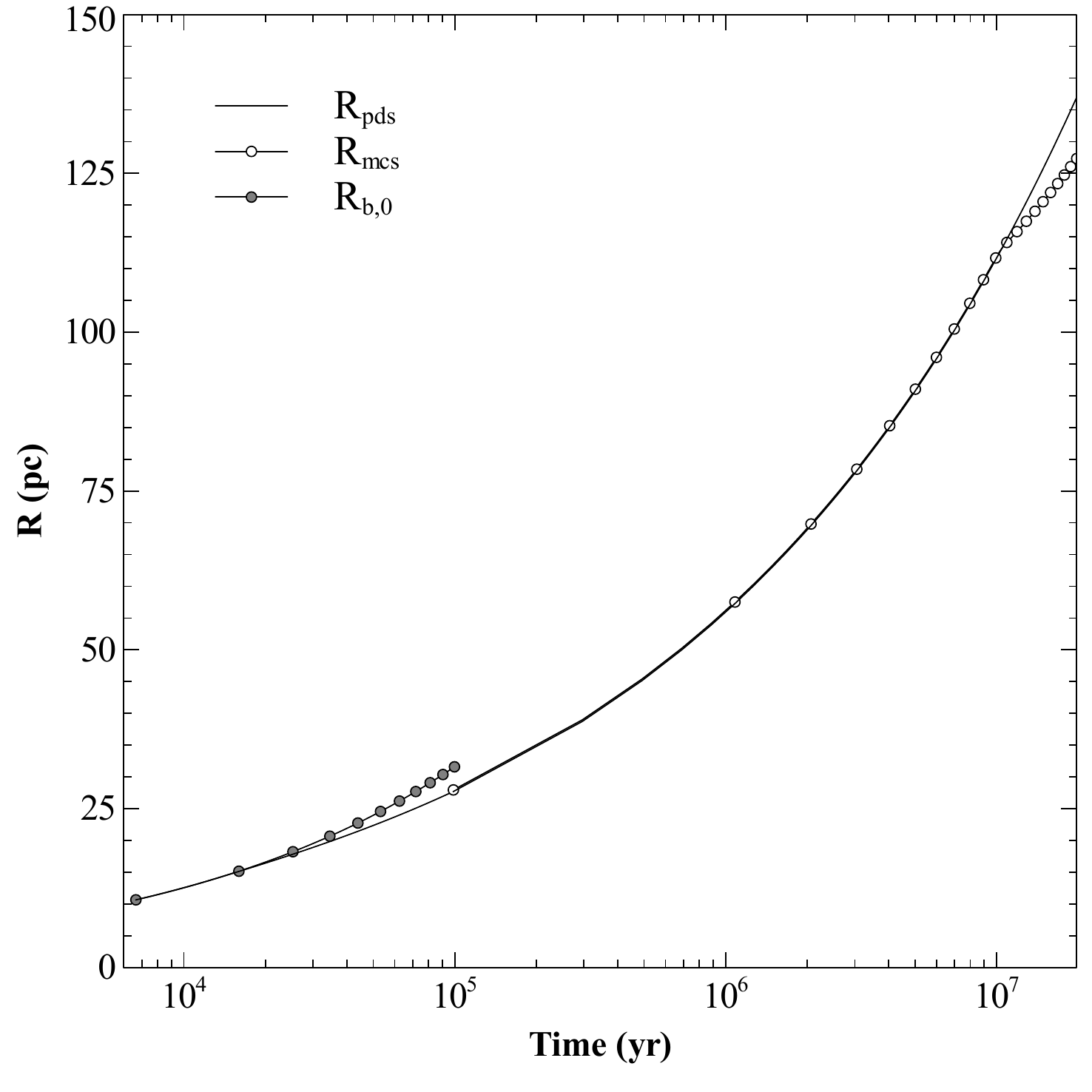}{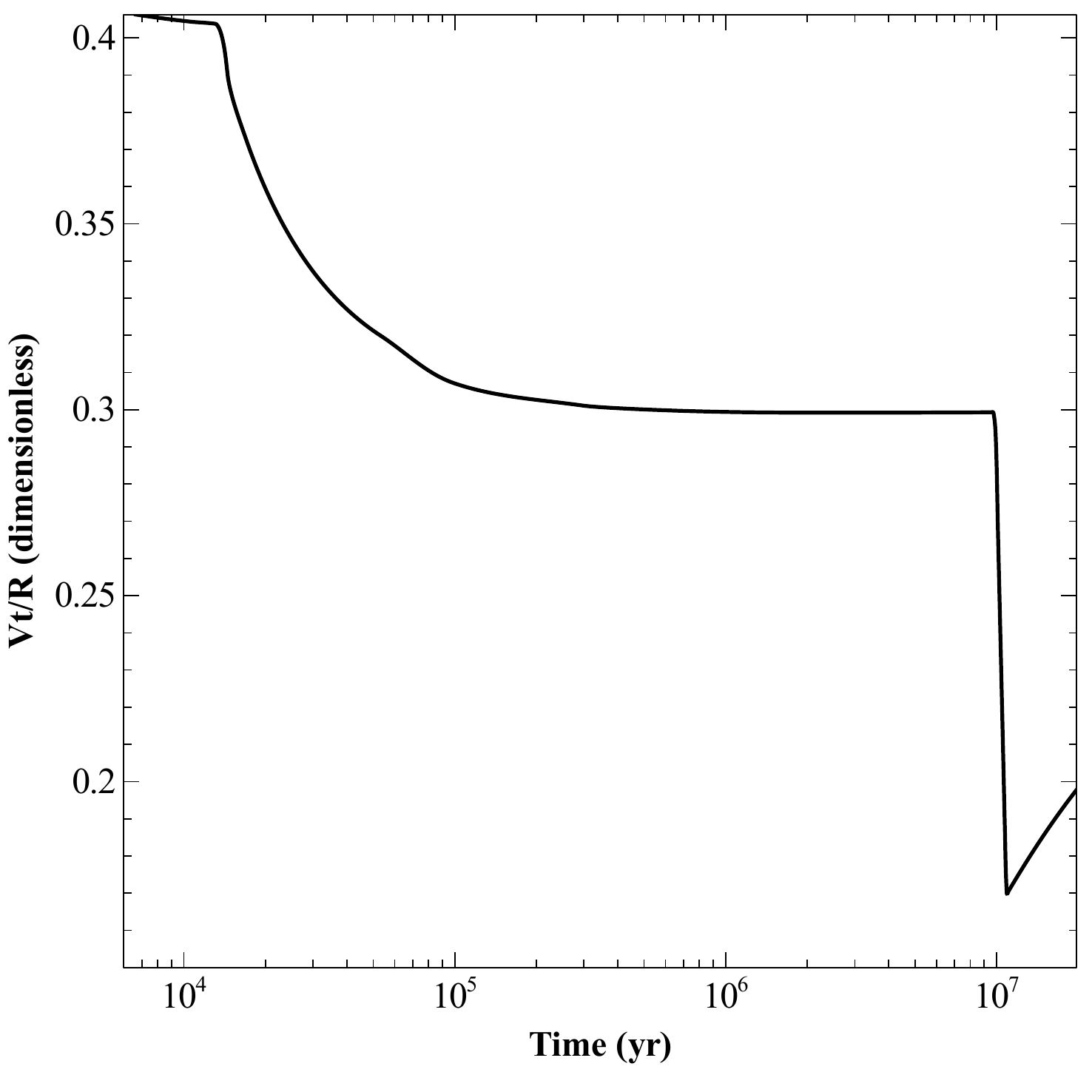}
\caption{
Left panel: Late-time evolution of blast-wave radius including transitions from ST to PDS and PDS to MCS phases. The line $R_{b,0}$ shows the ST evolution (up to $10^5$ yr) without including transition to PDS phase; the line $R_{pds}$ shows the PDS evolution without including transition to MCS phase; the line $R_{mcs}$ includes the full evolution.
Right panel: the deceleration index $m=Vt/R$ vs. time. 
\label{figPDSMCS}}
\end{figure}

\begin{figure}
\plottwo{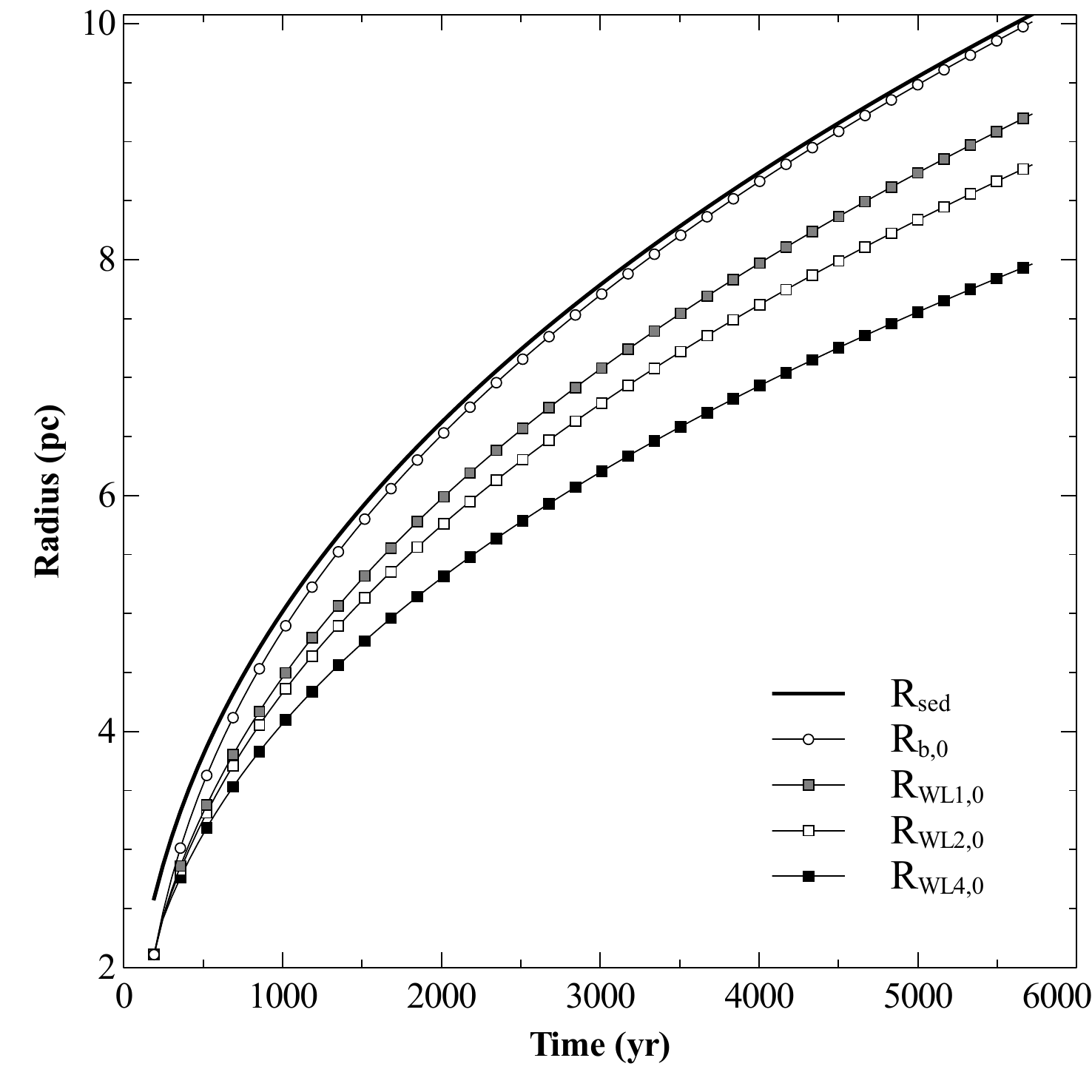}{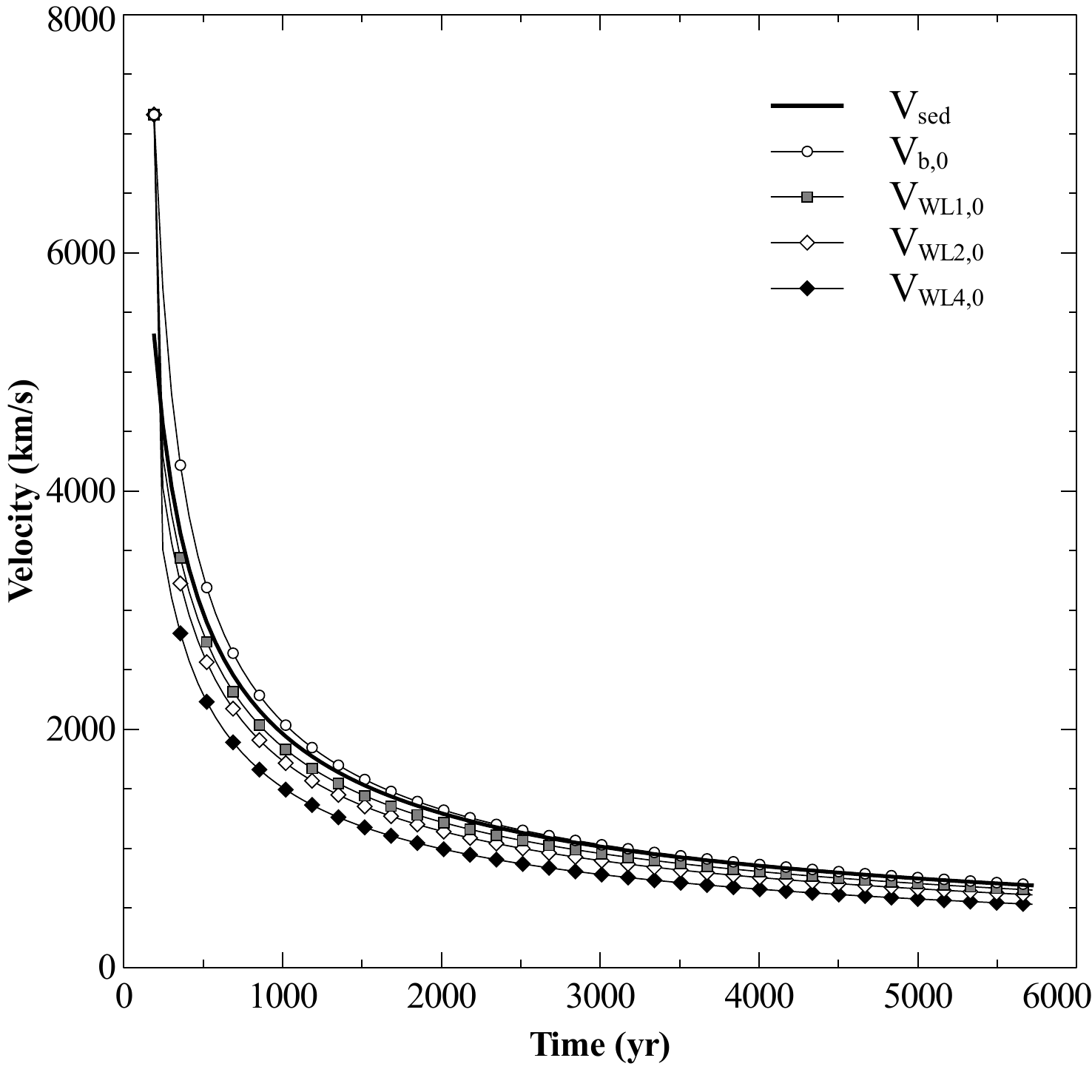}
\caption{
Comparison of SNR evolution in uniform ISM with SNR evolution in cloudy ISM . Left panel: comparison of  of blast-wave radius for pure Sedov-Taylor ($R_{sed}$) 
and n=0 ST model ($R_b,0$) with cloudy ISM models with $C/\tau$= 1, 2 and 4, matched to the n=0 ST model ($R_{WL1,0}$, $R_{WL2,0}$ and $R_{WL4,0}$). 
 The pure Sedov-Taylor solution and n=0 ST solutions have larger radius at all times than the solutions with cloudy ISM.
 Right panel: comparison of shock velocities for the same cases. The cloudy ISM models have shock velocity lower than the pure Sedov solution 
except at the very earliest times.
\label{figCloudy}}
\end{figure}


\begin{figure}
\plottwo{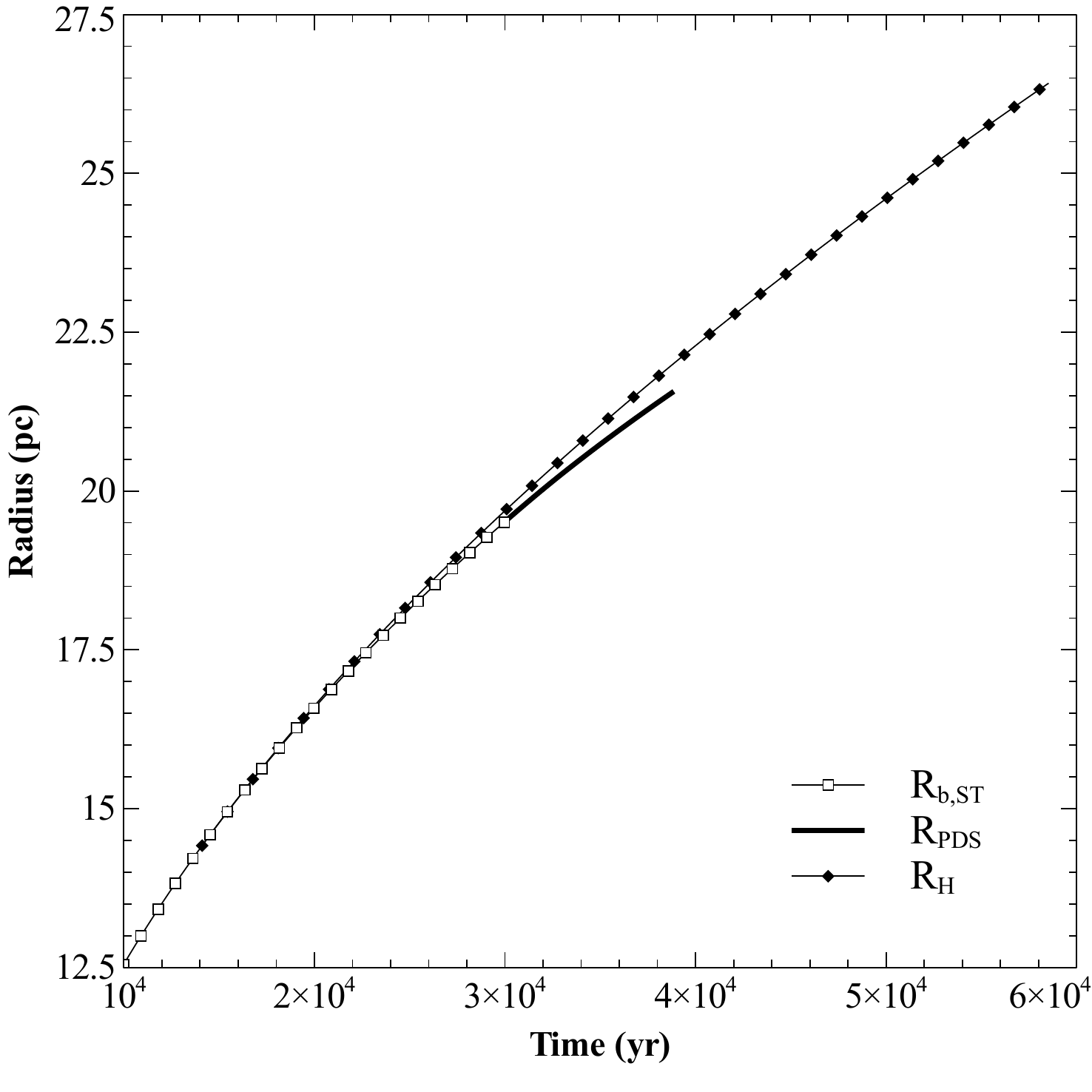}{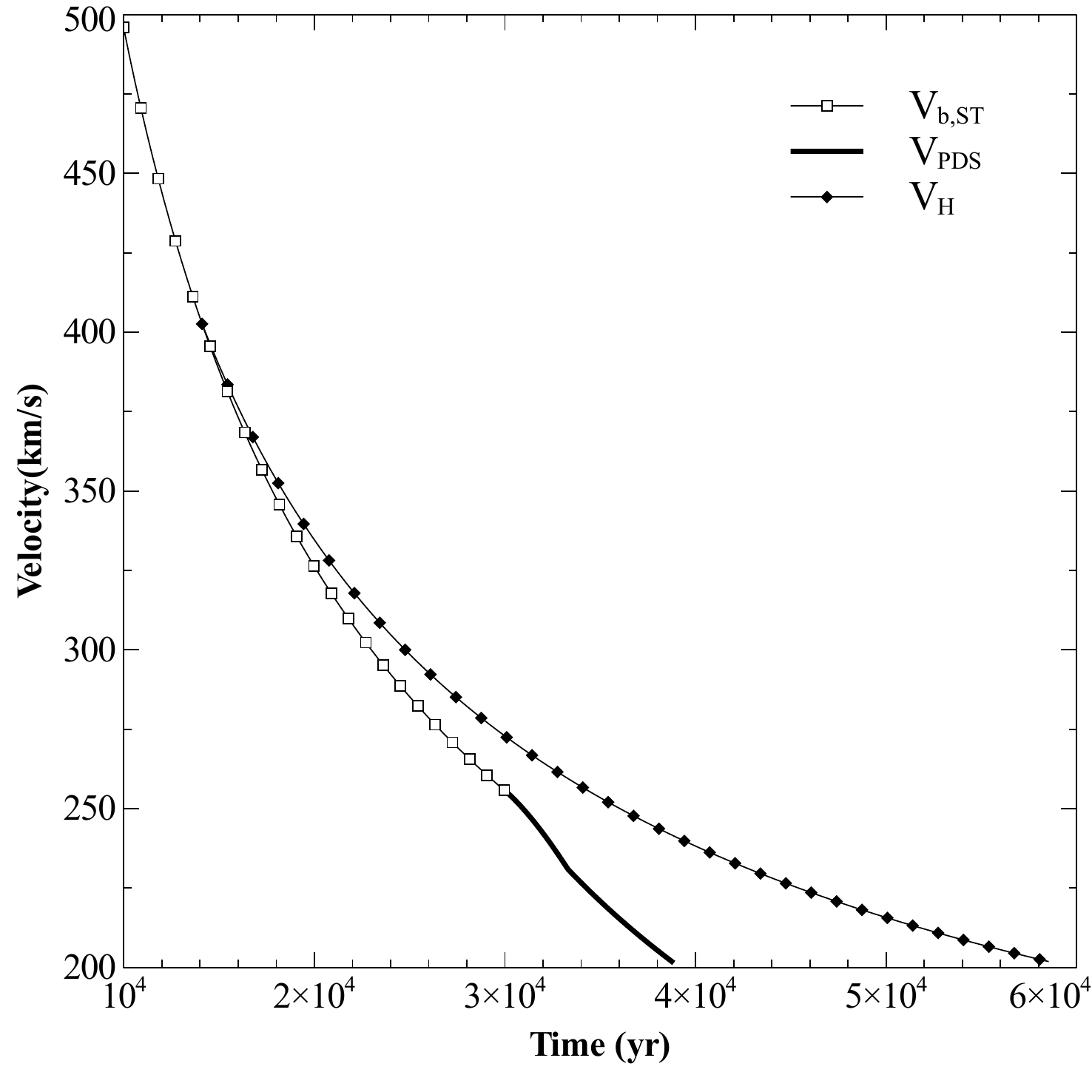}
\caption{
Left panel: Late-time evolution of blast-wave radius in a hot ISM (T=$10^6$K). The lines $R_{b,ST}$ and $R_{PDS}$ shows the standard
evolution in ST and PDS phases up to merger at $3.87\times10^4$yr. 
The line $R_{H}$ shows the evolution including swept up thermal energy from the hot ISM. In the latter case
merger occurs much later: at  $5.85\times10^4$yr.
Right panel: the shock velocity for the standard evolution ($V_{b,ST}$ and $R_{PDS}$), and for the hot ISM model ($V_{H}$). Both models
end with merger into the ISM at the same velocity.
\label{fighotISM}}
\end{figure}






\clearpage

\begin{deluxetable}{ccrrrrrrr}
\tabletypesize{\scriptsize}
\tablecaption{Dimensionless Emission Measures and Temperatures}
\tablewidth{0pt}
\tablehead{
\colhead{phase} & \colhead{n }  & \colhead{$C/\tau$ } & \colhead {$dEM_f$} &  \colhead{$dT_f$ } &  \colhead{$dEM_r$ }
  & \colhead{$dT_r$ }  
 
}
\startdata
 Ejecta Dominated  & 7   &  n/a &   0.753 & 1.228 &  0.611 & 0.538 \\   
 Ejecta Dominated  & 12   &  n/a &   0.916 & 1.129 & 11.82 & 0.119 \\  
 Sedov      &  n/a   &  n/a &   0.516 & 1.289 &  n/a &  n/a \\  
   WL      &  n/a  & 1 &  0.774 & 1.370 &  n/a &  n/a \\  
   WL      &  n/a  & 2 &  1.609 & 1.369 &  n/a &  n/a \\
   WL      &  n/a  & 4 &  6.932 & 1.383 &  n/a &  n/a \\
\enddata
\end{deluxetable}



\begin{thebibliography}{}

\bibitem[Borkowski et al.(2001)]{2001Bork} Borkowski, K.~J., Lyerly, W.~J., \& Reynolds, S.~P.\ 2001, \apj, 548, 820

\bibitem[Chevalier(1982)]{1982Chev} Chevalier, R.~A.\ 1982, \apj, 258, 790 

\bibitem[Cioffi et al.(1988)]{1988cioffi} Cioffi, D.~F., McKee, C.~F., \& Bertschinger, E.\ 1988, \apj, 334, 252

\bibitem[Cox \& Anderson(1982)]{1982CA} Cox, D.~P., \& Anderson, P.~R.\ 1982, \apj, 253, 268 

\bibitem[Ghavamian et al.(2013)]{2013Ghav} Ghavamian, P., Schwartz, S.~J., Mitchell, J., Masters, A., \& Laming, J.~M.\ 2013, \ssr, 178, 633 

\bibitem[Grevesse \& Sauval(1998)]{1998GS} Grevesse, N., \& Sauval, A.~J.\ 1998, \ssr, 85, 161 

\bibitem[Hamilton et al.(1983)]{1983Ham} Hamilton, A.~J.~S., Sarazin, C.~L., \& Chevalier, R.~A.\ 1983, \apjs, 51, 115 
\bibitem[Itoh(1978)]{1978I} Itoh, H.\ 1978, \pasj, 30, 489

\bibitem[Itoh et al.(2002)]{2002NCimB.117..359I} Itoh, N., Kawana, Y., \& Nozawa, S.\ 2002, Nuovo Cimento B Serie, 117, 359 


\bibitem[Leahy(2016)]{2016Leahy} Leahy, D.~A.\ 2016, arXiv:1612.00468 

\bibitem[Liang \& Keilty(2000)]{2000LK} Liang, E., \& Keilty, K.\ 2000, \apj, 533, 890 


\bibitem[Raymond et al.(1976)]{1976Ray} Raymond, J.~C., Cox, D.~P., \& Smith, B.~W.\ 1976, \apj, 204, 290


\bibitem[Russell \& Dopita(1992)]{RD1992} Russell, S.~C., \& Dopita, M.~A.\ 1992, \apj, 384, 508 

\bibitem[Tang \& Wang(2005)]{2005TW} Tang, S., \& Wang, Q.~D.\ 2005, \apj, 628, 205 

\bibitem[Truelove \& McKee(1999)]{1999truelove} Truelove, J.~K., \& McKee, C.~F.\ 1999, \apjs, 120, 299 

\bibitem[Vink(2012)]{2012Vink} Vink, J.\ 2012, \aapr, 20, 49 

\bibitem[White \& Long(1991)]{1991WL} White, R.~L., \& Long, K.~S.\ 1991, \apj, 373, 543 


\end{thebibliography}
\end{document}